\documentclass[a4paper,11pt,preprint]{article}

\usepackage{jcappub}
\usepackage[utf8]{inputenc}
\usepackage{cancel}
\usepackage{float}
\usepackage{graphicx}
\usepackage{hyperref}
\usepackage{natbib}
\usepackage{latexsym}
\usepackage{amsmath}
\usepackage{amssymb}
\usepackage{bbm}

\usepackage[normalem]{ulem}
\usepackage{pdfsync}
\usepackage{epsfig}
\usepackage{epstopdf}
\usepackage{subfigure}
\usepackage{color}
\usepackage{comment}
\usepackage{slashed}

\newcommand{\bmp}{\noindent\begin{minipage}{16cm}}
\newcommand{\emp}{\end{minipage}\vskip 7mm} 
\def\lsim{\mathrel{\raise.3ex\hbox{$<$\kern-.75em\lower1ex\hbox{$\sim$}}}}
\def\gsim{\mathrel{\raise.3ex\hbox{$>$\kern-.75em\lower1ex\hbox{$\sim$}}}}

\newcommand{\intron}[1]{}

\newcommand{\MeV}{\mathinner{\mathrm{MeV}}}
\newcommand{\GeV}{\mathinner{\mathrm{GeV}}}

\title{Gravitational wave constraints on the observable inflation}

\author{Erwin H. Tanin}
\author{and Tommi Tenkanen}

\affiliation{Department of Physics and Astronomy, Johns Hopkins University, \\
3400 N. Charles St., Baltimore, MD 21218, USA}
\emailAdd{etanin1@jhu.edu}
\emailAdd{ttenkan1@jhu.edu}

\abstract{Gravitational waves (GW) produced in the early Universe contribute to the number of relativistic degrees of freedom, $N_{\rm eff}$, during Big Bang Nucleosynthesis (BBN). By using the constraints on $N_{\rm eff}$, we present a new bound on how much the Universe could have expanded between horizon exit of the largest observable scales today and the end of inflation. We discuss the implications on inflationary models and show how the new constraints affect model selection. We also discuss the sensitivities of the current and planned GW observatories such as LIGO and LISA, and show that the constraints they could impose are always less stringent than the BBN bound.
}

\begin{document}
\maketitle

%

\section{Introduction}
\label{sec:intro}

Cosmic inflation, an early period of accelerated expansion, is the current paradigm for explaining the origins and properties of temperature fluctuations in the Cosmic Microwave Background radiation (CMB) and the large scale structure of the Universe \cite{Martin:2013tda,Chowdhury:2019otk}. Cosmic inflation is also successful in explaining why the Universe is spatially flat, homogeneous, and isotropic to a high degree, and why the abundance of topological defects predicted by grand unified theories of particle physics within our observable Universe is unobservably low. 

Among the parameters that are relevant to the curvature perturbations that seed the large scale structure formation, two have already been measured to a high precision: the magnitude of scalar perturbations at large scales (the amplitude $\mathcal{P}_\zeta(k_*) \simeq 2.1\times 10^{-9}$ at the reference scale $k_*=0.05\,{\rm Mpc}^{-1}$, corresponding to CMB temperature fluctuations $\delta T/T \sim 10^{-5}$) and how the perturbations change with scale (the spectral tilt, $n_s \equiv 1+{\rm d}\ln\mathcal{P}_\zeta/{\rm d}\ln k \simeq 0.965$), as measured by the Planck satellite \cite{Akrami:2018odb,Aghanim:2018eyx}. The Planck and BICEP2/Keck Array collaborations have also placed strong constraints on the magnitude of primordial gravitational waves (GWs), usually expressed in terms of the tensor-to-scalar ratio: $r\equiv \mathcal{P}_t(k_*)/\mathcal{P}_\zeta(k_*) < 0.06$, where $\mathcal{P}_t(k)$ represents the tensor power spectrum \cite{Ade:2018gkx}.

Yet, the amount of \textit{observable} inflation occurring between the horizon exit of the cosmologically relevant scales today and the end of inflation is unknown. This uncertainty makes it difficult to link a given model of inflation with the properties of observable cosmological perturbations that it supposedly explains (see e.g. Ref. \cite{Martin:2013tda}). The amount of observable inflation is usually characterized by the number of e-folds $N(k)\equiv \ln(a_{\rm inf}/a_k)$ between the scale factor $a_k=k/H_{\rm inf}$ at which the mode $k$ of interest exited the inflationary horizon $(a_{\rm inf}H_{\rm inf})^{-1}$ and that at the end of inflation $a_{\rm inf}$. Of particular interest is the number of e-folds $N(k=a_0H_0)$ corresponding to the size of our observable universe, as it is the total amount of inflation that we can conceivably observe\footnote{For convenience, by the currently observable Universe (the current ``horizon") we refer to the distance scale $1/a_0H_0$, where $a_0$ and $H_0$ are the scale factor and the Hubble parameter today.}. Thus, for concreteness, we choose to present some of our key results in terms of $N(k=a_0H_0)$. While the value $N(k=a_0H_0)\sim 60$ is usually assumed, the actual amount can differ considerably from this. Assuming that the post-inflationary expansion of the Universe is controlled by a set of perfect fluids, such as radiation and cold dark matter, transitions between different epochs are prompt, and the Hubble rate did not decrease much during inflation\footnote{Because the first slow-roll parameter is $\epsilon=\dot{H}/H^2\ll 1$, it is usually a very good approximation that the Hubble rate did not decrease much during inflation for the range of e-folds that is of interest here. In particular, this is the case for plateau models that give the best fit to the CMB data \cite{Akrami:2018odb}, such as the well-motivated Higgs \cite{Bezrukov:2007ep,Bauer:2008zj} or Starobinsky inflation \cite{Starobinsky:1980te} models, and therefore in this paper we maintain this assumption throughout the paper.}, the available range for the number of e-folds is $N(k=a_0H_0) = 18 -77$ \cite{Liddle:2003as,Dodelson:2003vq}, which in terms of the scale factor corresponds to inflationary expansion of the currently observable Universe by a factor $a_{\rm inf}/a_{k=a_0H_0} = 10^8 - 10^{33}$. Further constraints on reheating, and therefore on $N$, exist if one assumes certain inflationary potentials \cite{Martin:2010kz,Martin:2014nya}, and/or restricts the range of possible values of the equation of state parameter during reheating \cite{Munoz:2014eqa, Cook:2015vqa}. It should be emphasized, however, that by how much the Universe expanded between the horizon exit of scales that reside outside our current horizon and the end of inflation is something we cannot, unfortunately, answer with confidence. See, however, Ref. \cite{Remmen:2014mia} for model-dependent discussion on this aspect.

In this paper we utilize the constraints on the number of relativistic degrees of freedom, $N_{\rm eff}$, during Big Bang Nucleosynthesis (BBN) to derive a new bound on how much the Universe expanded between horizon exit of the cosmologically relevant scales today and the end of inflation. The new bound is valid within the standard assumptions of the inflationary dynamics and post-inflationary expansion history. We discuss implications on inflationary models and show how the new constraints affect model selection, and also the sensitivities of the current and planned gravitational wave observatories such as LIGO, LISA, Einstein Telescope, and BBO, and show that the constraints they could impose are always less stringent than the BBN bound.

The paper is organized as follows: in Sec. \ref{appendix:efolds}, we review the standard calculation of the number of e-folds between the horizon exit of a scale $k$ and the end of inflation, and present our main results in Sec. \ref{sec:GWs}. Sections \ref{sec:inflation} and \ref{sec:future} are devoted for discussion on inflationary models and future experiments, respectively. Finally, in Sec. \ref{sec:conclusions}, we conclude.

\section{The number of e-folds
}
\label{appendix:efolds}

We begin by presenting the number of e-folds between the horizon exit of a scale $k$ and the end of inflation. We mostly follow Ref. \cite{Liddle:2003as}, although we treat some steps somewhat differently and therefore choose to present our derivation of the otherwise well-known result in full. 

A scale $k$ is related to the present Hubble scale $H_0$ as
\begin{equation}
\frac{k}{a_0 H_0} = \frac{a_k H_{\rm inf}}{a_0 H_0} = e^{-N(k)}\frac{a_{\rm inf}}{a_{\rm RD}}\frac{a_{\rm RD}}{a_0}\frac{H_{\rm inf}}{H_0}\,,
\end{equation}
where $H_{\rm inf}$ is the Hubble parameter during inflation and $a_{\rm RD}$ is the scale factor at the time when the radiation-dominated epoch (RD) began after inflation, that is, we are not assuming that the Universe entered into the usual Hot Big Bang (HBB) epoch immediately after inflation but allow for an intermediary period of non-radiation dominated expansion between inflation and HBB. This gives the number of e-folds as
\begin{equation}
\label{N}
N(k)= \ln\left(\frac{a_{\rm inf}}{a_{\rm RD}} \right) +  \ln\left(\frac{a_{\rm RD}}{a_0} \right)  -  \ln\left(\frac{k}{a_0 H_{\rm inf}} \right)\,,
\end{equation}
which shows that the amount of expansion $N(k)$ between horizon exit of a scale $k$ and the end of inflation is completely determined by the post-inflationary expansion history. The question we would like to ask then is: given all observational constraints, what is the maximum value of $N(k)$ for a given $k$?

To answer this question, we have to find an expression for $N(k)$ in terms of observables and quantities one can hope to be able to compute from the underlying particle physics theory, such as the energy scale when the RD commenced after inflation. First, we assume that between inflation and RD the total energy density scaled (on average) as 
\begin{equation}
\rho(a) = \rho(a_{\rm inf})\left(\frac{a_{\rm inf}}{a}\right)^{3(1+w)}\,, 
\end{equation}
where $w\equiv p/\rho$ is the effective equation of state (EoS) parameter of the dominant fluid which is characterized by its energy density $\rho$ and pressure $p$, and the scaling of $\rho$ in terms of $a$ follows in the usual way from the continuity equation $\dot{\rho} = -3H(1+w)\rho$, where $H=\dot{a}/a$ and the overdot denotes derivative with respect to cosmic time $t$. Therefore, $w=0$ and $w=1/3$ correspond to an effectively matter-dominated post-inflation pre-RD epoch and an instant reheating into RD, respectively, whereas scenarios with $1/3<w<1$ are encountered in models where the total energy density of the Universe after inflation is dominated by the kinetic energy of a scalar field, either through oscillations in a steep potential (e.g. $V(\phi) \propto \phi^p$ with $p>4$), or by an abrupt drop in the potential \cite{Turner}. This is the case in e.g. quintessential inflation \cite{Peebles:1998qn}, where the inflaton field makes a transition from potential energy domination to kinetic energy domination at the end of inflation, reaching values of $w$ close to unity. The bound $w\leq 1$ comes from the requirement that the adiabatic sound speed of the dominant fluid does not exceed the speed of light. On the other hand, for $w>-1/3$ the Universe does not inflate. A plausible range for post-inflationary EoS parameter is therefore between these two values, and in the following we will maintain the dependence on $w$ explicitly in our calculations. Therefore, for the first term in Eq. (\ref{N}), we obtain
\begin{equation}
\label{1stterm}
\ln\left(\frac{a_{\rm inf}}{a_{\rm RD}}\right) = \frac{1}{3(1+w)}\ln\left(\frac{\rho_{\rm RD}}{\rho_{\rm inf}}\right)\,, 
\end{equation}
where $\rho_{\rm inf}\equiv\rho(a_{\rm inf})$ and $\rho_{\rm RD}\equiv\rho(a_{\rm RD})$ is the radiation energy density at the time the RD epoch began.

Assuming entropy conservation between RD and the present day and that the Universe thermalized quickly at the start of RD\footnote{This is a safe assumption, as the Standard Model plasma generically thermalizes in much less than one e-fold from its production, see e.g. Ref. \cite{McDonough:2020tqq}.}, we can write
\begin{equation}
    \frac{a_{\rm RD}}{a_0}\simeq
    \left(\frac{\pi^2}{30} \right)^{1/4}\frac{g_*^{1/3}(a_0)}{g_*^{1/12}(a_{\rm RD})}\frac{T_0}{\rho_{\rm RD}^{1/4}}\,,
\end{equation}
where $g_*$ is the number of effective relativistic degrees of freedom (assumed to be the same for entropy and energy density),
and $T_0 = 2.725$ K is the present-day CMB temperature \cite{Aghanim:2018eyx}. The second term in Eq. (\ref{N}) thus becomes
\begin{equation}
\label{2ndterm}
\ln\left(\frac{a_{\rm RD}}{a_0} \right) \simeq -66.1 -\ln\left(\frac{\rho_{\rm RD}^{1/4}}{10^{16}\,{\rm GeV}}\right)\,,
\end{equation}
where we have taken $g_*(a_0)=3.909$ and $g_*(a_{\rm RD})=106.75$. If $g_*(a_{\rm RD})$ was e.g. an order of magnitude larger or smaller, the first term above would change by only $O(0.1)$, and so we will henceforth neglect the $g_*(a_{\rm RD})$ dependence.

Next, we can express the tensor-to-scalar ratio $r$ in the slow-roll approximation (see e.g. Ref. \cite{Baumann:2009ds}) as
\begin{equation}
\label{rDefinition}
r = \frac{8}{M_{\rm P}^2 \mathcal{P}_\zeta(k_*)}\left(\frac{H_{\rm inf}}{2\pi}\right)^2\,,
\end{equation}
where $M_{\rm P}$ is the reduced Planck mass and we assumed that the Hubble scale and the amplitude of perturbations did not change between the horizon exit of the scale $k$ and the pivot scale $k_*$ where $r$ is measured (usually $k_* = 0.05\,{\rm Mpc}^{-1}$, see e.g. Ref. \cite{Ade:2018gkx}). With this, the last term in Eq. (\ref{N}) becomes
\begin{equation}
\label{3rdterm}
-\ln\left(\frac{k}{a_0 H_{\rm inf}}\right) \simeq 128.3 + \frac12\ln\left(\frac{r}{0.1}\right)
- \ln\left(\frac{k}{a_0 H_0}\right)\,,
\end{equation}
where we have used the measured value of $\mathcal{P}_\zeta(k_*)$ and normalized $k$ to the present horizon $a_0 H_0$. We emphasize that up to this point we have not specified $k$ and that it can be freely chosen amongst all modes that exited the horizon during inflation, although usually only those that can be probed by CMB experiments are of astrophysical interest.

Finally, by assuming that the total energy density did not decrease much during the final $N$ e-folds so that
\begin{equation}
\label{rho_inf}
\rho_{\rm inf} = 3 H_{\rm inf}^2M_{\rm P}^2 \simeq (10^{16}\,{\rm GeV})^4\left(\frac{r}{0.1}\right)\,,
\end{equation}
as given by Eq. \eqref{rDefinition}, and substituting Eqs. \eqref{1stterm}, \eqref{2ndterm}, and \eqref{3rdterm} into Eq. \eqref{N}, we obtain the result
\begin{eqnarray}
    \label{efolds}
    N(k) \simeq  62+\frac{1+3w}{6(1+w)}\ln\left(\frac{r}{0.1}\right)+\frac{1-3w}{3(1+w)}\ln\left(\frac{\rho_{\rm RD}^{1/4}}{10^{16}\GeV}\right)- \ln\left(\frac{k}{a_0 H_0}\right)\,.
\end{eqnarray}
This is our final result for the number of e-folds. By assuming that before the usual Hot Big Bang epoch the Universe was effectively matter-dominated, i.e. by setting $w=0$, we recover the usual result discussed in e.g. Ref. \cite{Liddle:2003as}. 

We will now determine the maximum possible value of $N(k)$. From Eqs. \eqref{N}, \eqref{1stterm}, \eqref{2ndterm}, it is clear that for given $r$ and $\rho_{\rm RD}$ (smaller than $\rho_{\rm inf}$), $N$ is largest when the equation of state $w$ during the intermediary epoch is maximized, $w\approx 1$. Furthermore, if $w$ is stiff, i.e. $1/3<w<1$, then Eq.~\eqref{efolds} tells us that $N(k)$ is maximized when $\rho_{\rm RD}$ is minimized. At the very least, radiation domination must commence before the onset of BBN, so $\rho_{\rm RD}^{1/4}\gtrsim T_{\rm BBN}\sim 5\MeV$ \cite{Hasegawa:2019jsa} or, equivalently,
\begin{equation}
    \ln\left(\frac{\rho_{\rm RD}^{1/4}}{10^{16}\GeV}\right)\gtrsim -42\,. \label{TRDmin1}
\end{equation}
By setting $\rho_{\rm RD}$ in such a way that the above condition is saturated and $w\approx 1$, we find an upper bound on $N$:
\begin{equation}
    N(k)\lesssim 76+\frac{1}{3}\ln\left(\frac{r}{0.1}\right)-\ln\left(\frac{k}{a_0 H_0}\right)\,.\label{NTBBN}
\end{equation}
Thus, for the maximum allowed value of the tensor-to-scalar ratio, $r=0.06$, we find $N\lesssim 76$ for the largest observable scale $k=a_0 H_0$, in good agreement with the well-known result of Ref. \cite{Liddle:2003as}, where $\rho_{\rm RD}^{1/4}\sim 1\MeV$ was assumed. This is the maximum value of $N(k=a_0H_0)$ one can obtain within the standard assumptions discussed above. 

For clarity, we note that by allowing $H$ to evolve during inflation the above calculation does not hold but one may have to solve an implicit equation for $N$, depending on the model of inflation, see e.g. Refs.~\cite{Liddle:2003as,Dodelson:2003vq,Martin:2010kz,Martin:2013tda,Takahashi:2018brt}. In extreme cases, one can obtain a considerably higher value up to $N(k=a_0H_0)\sim 100$, although a number this large requires $H$ to change by a highly implausible amount during inflation \cite{Liddle:2003as}. However, as discussed in Sec. \ref{sec:intro}, in models that give the best fit to the CMB data this does not happen.

\section{Gravitational waves and $N_{\rm eff}$}
\label{sec:GWs}

While the results \eqref{efolds}, \eqref{NTBBN} are robust within our assumptions of the nature of inflation and reheating, they do not take into account the constraints the lack of observation of a stochastic GW background imposes on $w$ and $\rho_{\rm RD}$.

Let us therefore consider gravitational waves. During a stiff epoch, the energy density parameter $\Omega_{\rm GW}$ of gravitational waves gets amplified as the universe expands \cite{Giovannini:1998bp,Giovannini:1999bh}. The lower the RD scale $\rho_{\rm RD}$, the more e-folds the stiff epoch lasts, and hence the more amplification $\Omega_{\rm GW}$ receives. This means $\rho_{\rm RD}$ is not allowed to be too low since the BBN bound on the number of extra relativistic degrees of freedom imposes an upper limit on the amount of gravitational waves present during BBN.

In the presence of a stiff epoch between the end of inflation and the beginning of radiation domination, the present-day GW energy density spectrum\footnote{The GW energy density spectrum is defined as the GW energy density $\rho_{\rm GW}$ per unit logarithm of frequency normalized to the critical density $\rho_{\rm crit}\equiv 3M_{\rm P}^2H^2$, $\Omega_{\rm GW}\equiv ({\rm d}\rho_{\rm GW}/{\rm d}\text{ln}f)/\rho_{\rm crit}$.} $h^2\Omega_{\rm GW}^{(0)}(f)$ originated from inflation is enhanced relative to that in the absence of a stiff epoch
\begin{equation}
h^2\Omega_{\rm GW,\,plat.}^{(0)}\simeq 1\times 10^{-16}\left(\frac{r}{0.1}\right)\,, 
\end{equation}
as \cite{Figueroa:2019paj}
\begin{equation}
    h^2\Omega_{\rm GW}^{(0)}(f)\simeq C(w)h^2\Omega_{\rm GW,\,plat.}^{(0)} \left(\frac{f}{f_{\rm RD}}\right)^{2\left(\frac{3w-1}{3w+1}\right)}\,, \label{bluetilt}
\end{equation}
where the subscript ``plat." refers to "plateau" (no tilt), $f_{\rm RD}\equiv k_{\rm RD}/(2\pi a_0)$ is the present-day frequency of the mode $k_{\rm RD}\equiv a_{\rm RD}H_{\rm RD}$ that matches the horizon size at the onset of RD, the expression applies for $f\gg f_{\rm RD}$, and $C(w)$ is an $O(1)$ factor that depends on $w$ and how abruptly the universe transitions from stiff-fluid domination to RD. We will set, without losing much accuracy, $C(w)=1$ as it ranges from 1 to 1.3 (1.8) for instantaneous (smooth) transition. It should be noted that Eq.~\eqref{bluetilt} does not account for the slight red spectral tilt $n_{\rm t}\equiv {\rm d}\ln\mathcal{P}_t/{\rm d}\ln k$ of the tensor power spectrum expected in slow-roll inflationary scenarios, which has been constrained down to $-n_{\rm t}\lesssim 0.008$ at around the CMB pivot scale  \cite{Akrami:2018odb,Ade:2018gkx}. In the absence of running of the spectral index, the spectral tilt can reduce the GW energy density by at most a factor of $\left(e^{76}\right)^{0.008}\sim 1.84$, which in the end weakens our constraint on $N(k)$ by only $\Delta N(k)=O(0.1)$. We will therefore neglect this effect.

The constraint on the number of extra relativistic degrees of freedom $\Delta N_{\rm eff}\lesssim 0.2$ during BBN \cite{Cyburt:2015mya} sets an upper bound on the GW energy density today \cite{Caprini:2018mtu}
\begin{equation}
    \int_{f_{\rm BBN}}^{f_{\rm inf}}h^2\Omega_{\rm GW}^{(0)}(f){\rm d}(\ln f)<1\times 10^{-6}\,,\label{BBNbound}
\end{equation}
where $f_{\rm BBN}$ and $f_{\rm inf}$ are the present-day frequencies of the modes that match the horizon size at BBN and the end of inflation, respectively. Evaluating the integral above with the help of Eq. \eqref{bluetilt} and $f_{\rm inf}\gg f_{\rm BBN}$, we arrive at
\begin{equation}
    h^2\Omega_{\rm GW}^{(0)}(f_{\rm inf})\lesssim 2\times 10^{-6} \left(\frac{3w-1}{3w+1}\right)\,. \label{DeltaNeffGW}
\end{equation}
Next, using Eqs. \eqref{1stterm}, \eqref{rho_inf}, \eqref{bluetilt}, and
\begin{equation}
    \frac{f_{\rm inf}}{f_{\rm RD}}=\frac{a_{\rm inf}H_{\rm inf}}{a_{\rm RD}H_{\rm RD}}\simeq \left(\frac{a_{\rm inf}}{a_{\rm RD}}\right)^{-(3w+1)/2}\,,
\end{equation}
we can rewrite Eq. \eqref{DeltaNeffGW} as
\begin{equation}
    \ln\left(\frac{\rho_{\rm RD}^{1/4}}{10^{16}\GeV}\right)\gtrsim \Theta_{\rm BBN}(w,r)\,,\label{TrehBBN}
\end{equation}
with
\begin{eqnarray}
    \Theta_{\rm BBN}(w,r)\equiv -\frac{3(1+w)}{4(3w-1)}\left[24+\ln\left(\frac{3w-1}{3w+1}\right)\right]+\frac{3w+1}{2(3w-1)}\ln\left(\frac{r}{0.1}\right)\,. \label{ThetaBBN}
\end{eqnarray}

\begin{figure}
\centering
\includegraphics[scale=0.8]{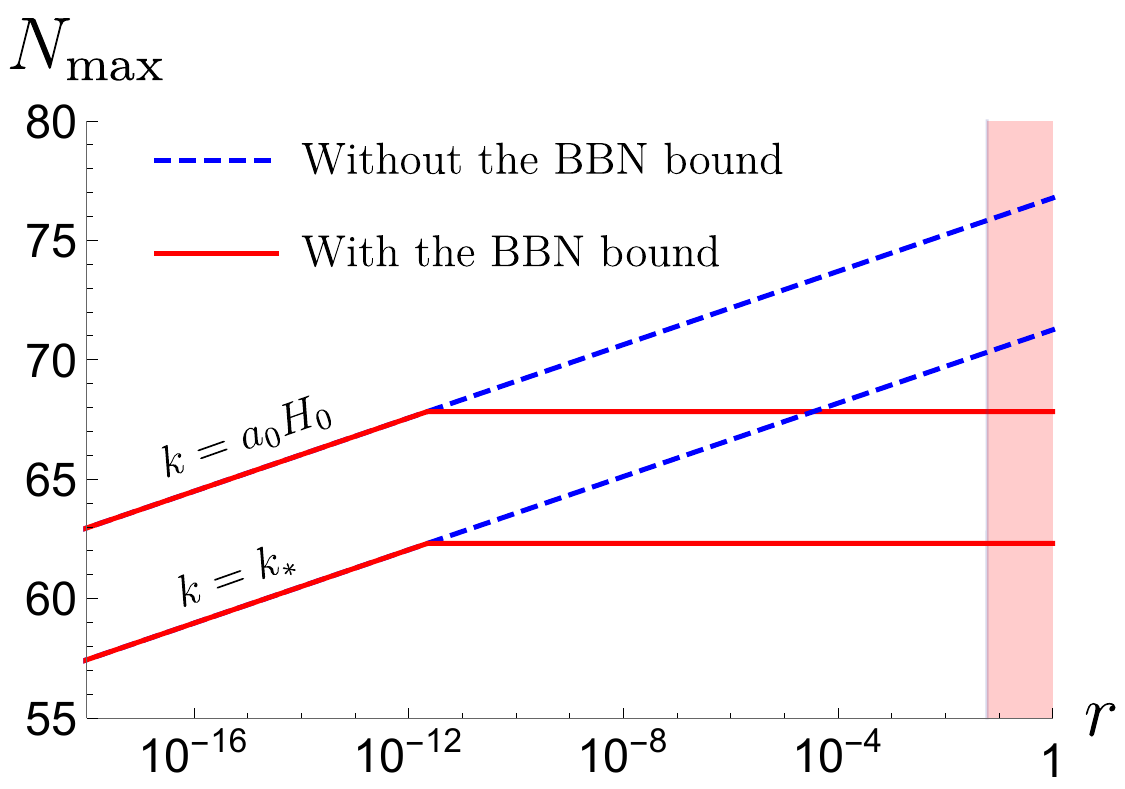}
\caption{Maximum allowed $N$ for the largest observable scale $k=a_0H_0=0.0002\text{ Mpc}^{-1}$ and the CMB pivot scale $k=k_*=0.05\text{ Mpc}^{-1}$ before and after imposing the BBN bound, Eq.~\eqref{BBNbound}. The shaded region has been ruled out by the non-observation of primordial B-mode polarization on the CMB \cite{Ade:2018gkx}.}
\label{fig:Nmax}
\end{figure}

We thus need
\begin{equation}
    \ln\left(\frac{\rho_{\rm RD}^{1/4}}{10^{16}\GeV}\right)\gtrsim \text{max}\left[\Theta_{\rm BBN}(w,r),-42\right]\,, \label{TRDmin2}
\end{equation}
where we have included also the previous bound from Eq. \eqref{TRDmin1}. If $r\lesssim 10^{-12}$, it is always the case that $\Theta_{\rm BBN}\lesssim -42$, meaning that the condition \eqref{TRDmin2} reduces to \eqref{TRDmin1} and the upper limit \eqref{NTBBN} on $N(k=a_0H_0)$ remains applicable. On the other hand, if $r\gtrsim 10^{-12}$, there are values of $w$, including $w=1$, for which \eqref{TRDmin2} is a stricter constraint than \eqref{TRDmin1}. However, it turns out that $N(k)$ remains to be maximized at $w=1$. Substituting $w=1$ and the value of $\rho_{\rm RD}$ that saturates \eqref{TRDmin2} into \eqref{efolds}, we find
\begin{equation}
    N(k)\lesssim N_{\rm max}=68-\ln\left(\frac{k}{a_0 H_0}\right)\,, \label{NmaxBBN}
\end{equation}
independently of the value of $r$. This is a new bound and our most important result. As can be seen in Figure~\ref{fig:Nmax}, for the maximum allowed value of tensor-to-scalar ratio $r\simeq 0.1$, the new bound is more stringent by $\Delta N \simeq -8$, which corresponds to a $3\times 10^{-4}$ reduction in the maximum $a_{\rm inf}/a_{k}$, i.e. the new bound places a constraint on the amount of inflationary expansion between the horizon exit of the cosmological scales and the end of inflation which is roughly four orders of magnitude more stringent than the previous bound when measured in terms of growth in the scale factor.

\pagebreak
\section{Constraints on inflationary models}
\label{sec:inflation}

Let us then discuss constraints on inflationary models by considering the very general action 
\begin{equation}
\label{jordanframe}
S_J = \int d^4x \sqrt{-g}\left[\frac12F(R(g_{\mu\nu},\Gamma),\phi) - \frac{1}{2}K(\phi) g^{\mu\nu}\nabla_{\mu}\phi\nabla_{\nu}\phi - V(\phi) \right]\,,
\end{equation}
where $g_{\mu\nu}$ and $\Gamma$ are the space-time metric and connection, respectively, $R$ is the Ricci scalar, $K(\phi)$ is a non-singular function of the scalar field $\phi$ (which we call the inflaton field), and $V(\phi)$ is the inflaton potential. Assuming slow-roll, the inflationary dynamics are characterized by the usual slow-roll parameters
\begin{equation}
\label{SRparameters1}
	\epsilon \equiv \frac{1}{2}M_{\rm P}^2 \left(\frac{V'}{V}\right)^2 \,, \quad
	\eta \equiv M_{\rm P}^2 \frac{V''}{V} \,,
\end{equation}
where the primes denote derivatives of the inflaton potential with respect to the field and in slow-roll $\epsilon\,, |\eta | \ll 1$, whereas the number of $e$-folds between the horizon exit of the scale where measurements are made (the pivot scale $k_*=0.05\, {\rm Mpc}^{-1}$) and the end of inflation is given by
\begin{equation}
\label{Ndef}
	N = \frac{1}{M_{\rm P}^2} \int_{\phi_{\rm end}}^{\phi_*} {\rm d}\phi \, V \left(\frac{{\rm d}V}{{\rm d} \phi}\right)^{-1}.
\end{equation}
The field value at the end of inflation, $\phi_{\rm end}$, is defined via $\epsilon(\phi_{\rm end})= 1$, and the field value at the time when the pivot scale exited the horizon is denoted by $\phi_*$. The leading order expression for the main observables -- the spectral tilt and tensor-to-scalar ratio -- are then given by
\begin{equation}
\label{nsralpha}
n_s -1 \simeq -6\epsilon + 2\eta\,, \quad
r \simeq 16\epsilon\,,
\end{equation}
which can be used to relate the number of required $e$-folds to the potential and other parameters of the model in the usual way.

\begin{figure}
\centering
\includegraphics[scale=0.60]{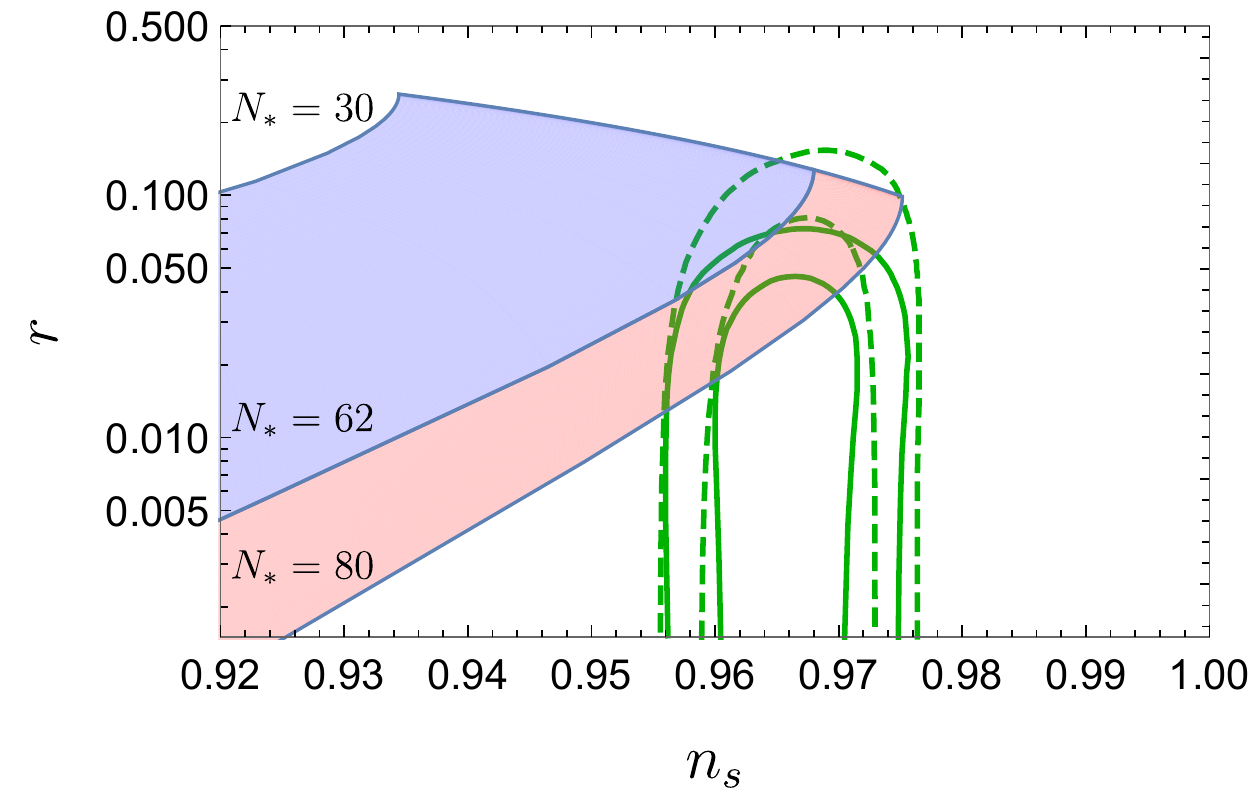}
\includegraphics[scale=0.60]{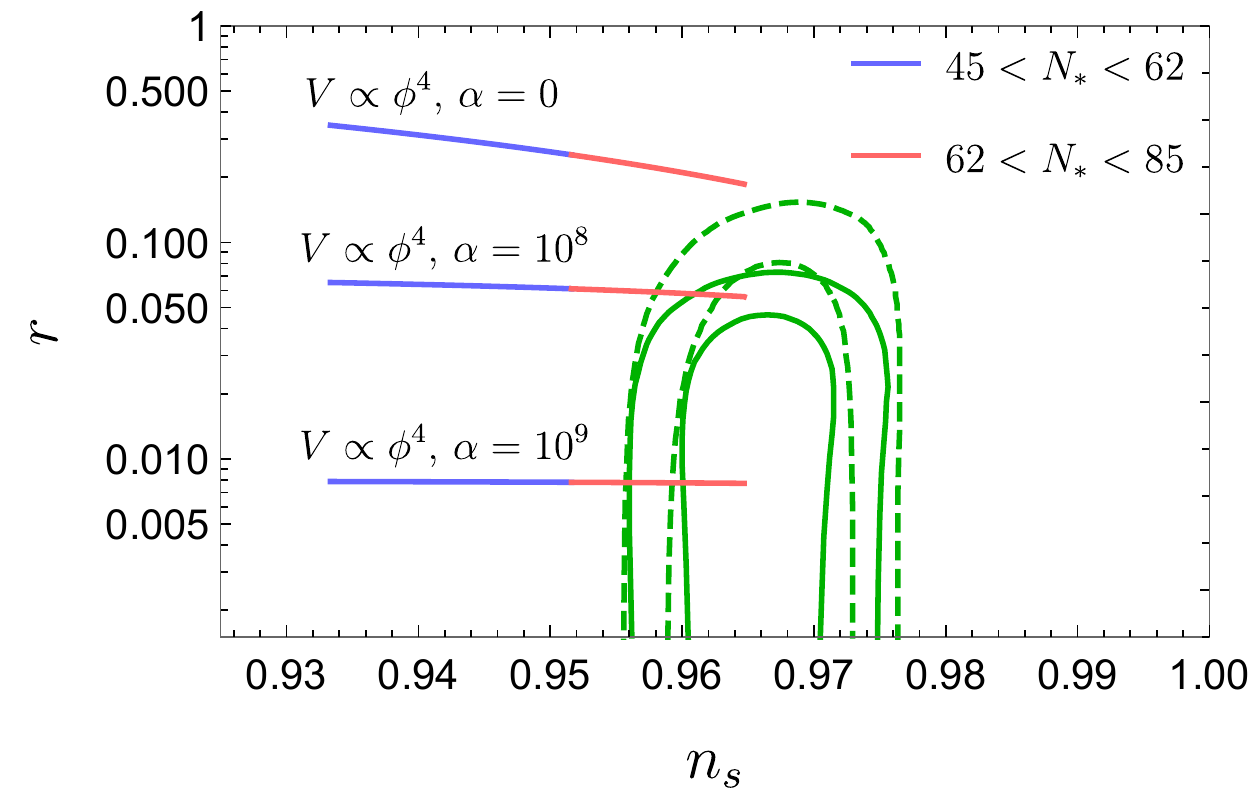}\\
\includegraphics[scale=0.60]{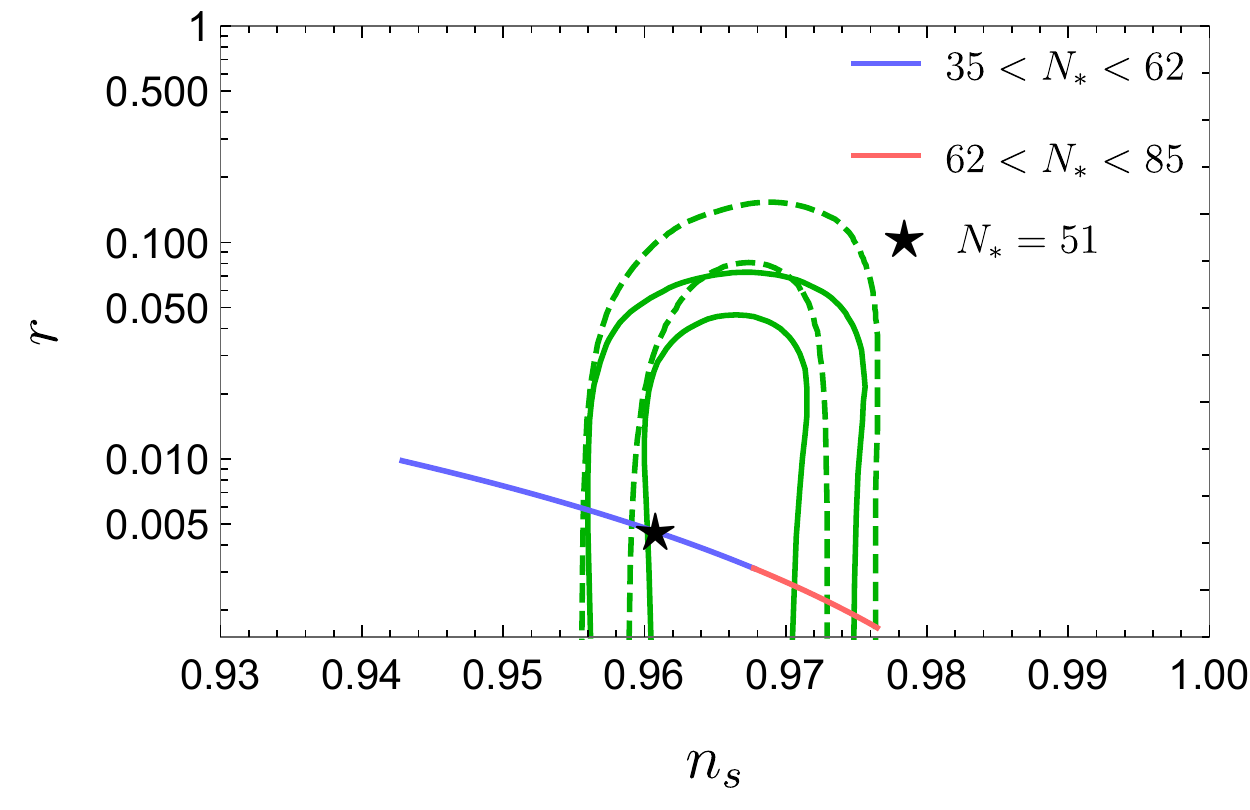}
\includegraphics[scale=0.60]{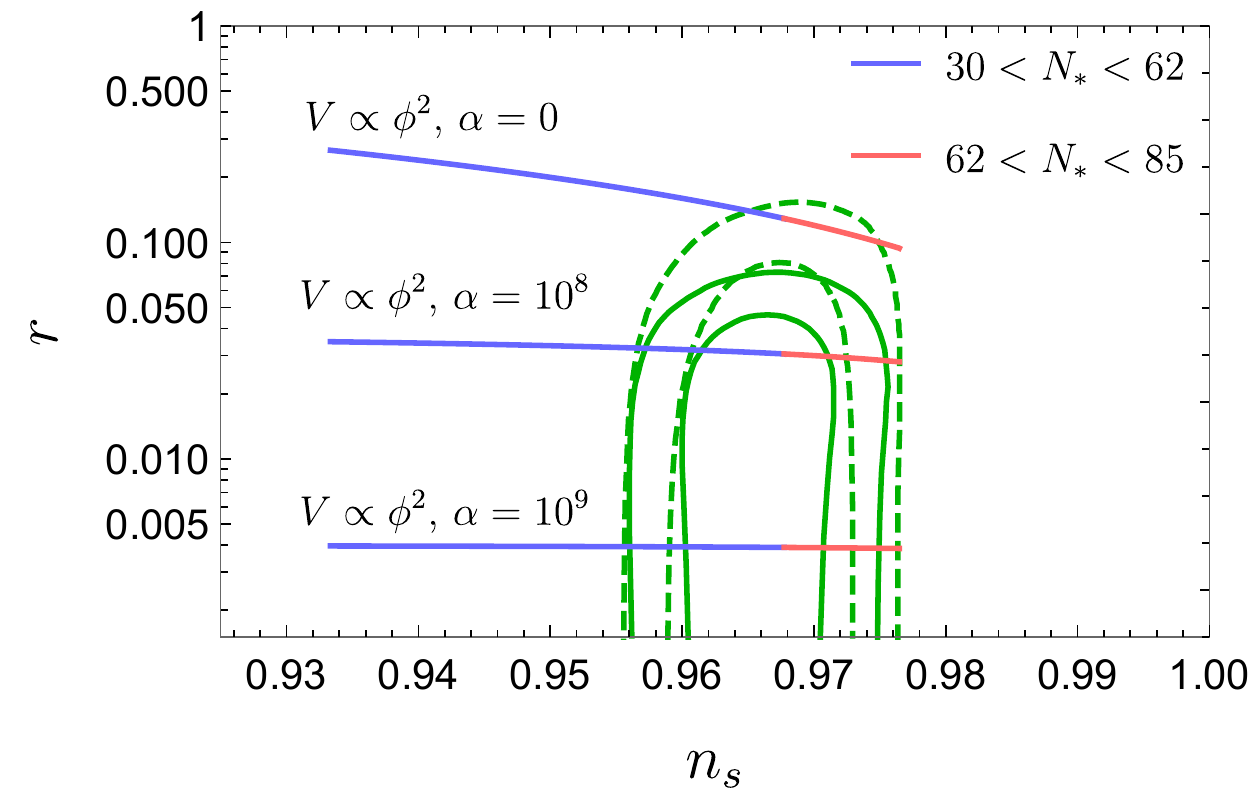}
\caption{A depiction of how the upper bound on the number of $e$-folds $N_*=N(k_*)<62$, Eq.~\eqref{NmaxBBN}, translates into constraints on the $(n_s,r)$ predictions of specific inflationary models: standard natural inflation (\textit{top-left}), quartic inflation with an $\alpha R^2$ term in Palatini gravity (\textit{top-right}), the usual Starobinsky inflation with metric gravity (\textit{bottom-left}), and quadratic inflation with an $\alpha R^2$ term in Palatini gravity (\textit{bottom-right}). The \textit{blue lines/regions} are within the GW constraints found in this paper while the \textit{red line/regions} are ruled out by them. The star symbol marks the predictions for $N(k_*)= 51$, which is the preferred value for Starobinsky inflation (see the main text). Also shown are the 68\% and 95\% CL contours from Planck TT+$\tau$ prior+lensing+BAO (\textit{green-dashed}) and Planck TT+$\tau$ prior+lensing+BAO+BICEP2/Keck (\textit{green-solid}) \cite{Ade:2018gkx}.}
\label{ns_r_fig}
\end{figure}

In the following, we will consider four different models to highlight the importance of our results, in particular the new bound \eqref{NmaxBBN}. First, we consider the well-known natural inflation model with a cosine potential
\begin{equation}
    V(\phi) = \Lambda^4\left[1+\cos\left(\frac{\phi}{f}\right) \right]\,,
\end{equation}
where $\Lambda$ and $f$ are mass scales determined by the underlying high energy theory \cite{Freese:1990rb} and the predictions for the spectral tilt and tensor-to-scalar ratio can be computed in the usual way from Eq. \eqref{nsralpha}. They are shown in Fig. \ref{ns_r_fig}. 

Second, we will consider the scenarios where in addition to the usual Einstein-Hilbert term $M_{\rm P}^2 R$ the action also contains an additional second-order term $\alpha R^2$, i.e. 
\begin{equation}
\label{Fterm}
    F(R(g_{\mu\nu},\Gamma),\phi) = M_{\rm P}^2 R(g_{\mu\nu},\Gamma) + \alpha R^2(g_{\mu\nu},\Gamma)\,,
\end{equation}
which includes the usual Starobinsky inflation model \cite{Starobinsky:1980te}, where inflation is driven by a scalar degree of freedom encapsulated in $R^2$. This model predicts (at lowest order in $1/N$)
\begin{equation}
    n_s(k_*) \simeq 1 - \frac{2}{N(k_*)} \quad r(k_*) \simeq \frac{12}{N^2(k_*)}\,,
\end{equation}
which fits very well to the data for $N(k_*)\gtrsim 45$, as shown in Fig. \ref{ns_r_fig}. In particular, this is the case for $N(k_*)= 51$, which is what studies on reheating in Starobinsky inflation suggest as the preferred value \cite{Gorbunov:2010bn,Bezrukov:2011gp}.

Finally, we will consider the action \eqref{Fterm} in two different cases where it is amended with $V(\phi) \propto \phi^n, n=2,4$ and where instead of the usual metric theory of gravity (with $\Gamma=\Gamma(g_{\mu\nu})$), we consider {\it Palatini} gravity where  $\Gamma=\Gamma(g_{\mu\nu},\phi)$ is a set of degrees of freedom which are a priori independent of the metric (see e.g. Refs. \cite{Bauer:2008zj,Tenkanen:2020dge}). This scenario has not only sparked interest in the community recently (see e.g. Refs. \cite{Enckell:2018hmo,Antoniadis:2018ywb,Karam:2018mft,Antoniadis:2018yfq,Tenkanen:2019jiq,Tenkanen:2019wsd,Tenkanen:2020cvw}) but is particularly interesting in our context, as for large enough $\alpha$ the addition of the $\alpha R^2$ term can resurrect models which are otherwise disfavored, in particular those where the inflaton potential is of simple polynomial type $V(\phi) \propto \phi^n$ -- given that the number of e-folds $N(k_*)$ was large enough. This is highlighted by the quadratic inflation model which as such is disfavored by data due to the large value it predicts for $r$. However, with the $\alpha R^2$ term in Palatini gravity the spectral index and tensor-to-scalar ratio become \cite{Tenkanen:2019wsd}
\begin{equation}
     n_s(k_*) \simeq 1 - \frac{2}{N(k_*)} \quad r(k_*) \simeq
     \frac{8}{N(k_*) + 96\pi^2P_\zeta(k_*) \alpha}\,,
\end{equation}
which for $\alpha\gtrsim 10^8\,, N(k_*)\gtrsim 45$ is compatible with data; see Fig. \ref{ns_r_fig}. On the other hand, for $V(\phi) \propto \phi^4$ one finds
\cite{Enckell:2018hmo}
\begin{equation}
    n_s(k_*) \simeq 1 - \frac{3}{N(k_*)}\,, \quad
    r(k_*) \simeq \frac{16}{N(k_*) + 96\pi^2P_\zeta(k_*) \alpha}\,.
\end{equation}
 As is evident from the expression for $n_s$, even for the largest allowed $N(k_*)$ given by Eq. \eqref{NmaxBBN}, the predicted value is disfavored by the most recent Planck data, see Fig. \ref{ns_r_fig}.

The results show that regardless of the actual number of e-folds, some interesting and well-motivated models (including the standard natural inflation) are clearly disfavored by data. While there are some caveats related to the way the Universe expanded after inflation (essentially whether the post-inflationary EoS parameter $w$ and/or the inflationary Hubble scale $H_{\rm inf}$ remained approximately constant or not), the assumptions we have made are fairly standard and well-motivated, and the above examples therefore highlight the importance of our results for inflationary models.

\section{Future experiments}
\label{sec:future}

Let us then discuss what are the prospects of future experiments for making the bound even more stringent. 

First, if the number of the extra relativistic degrees of freedom $N_{\rm eff}$ was measured to an accuracy better than the current bound $\Delta N_{\rm eff}\lesssim 0.2$, the maximum value \eqref{NmaxBBN} would go down by $1/4\ln\left(0.2/\Delta N_{\rm eff}\right)$. For instance, an order of magnitude improvement in the $\Delta N_{\rm eff}$ upper bound could be obtained from the constraints on the Hubble rate at the CMB decoupling~\cite{Smith:2006nka,Caprini:2018mtu}. These constraints were recently updated in Ref. \cite{Clarke:2020bil}, and while they were indeed found to be more stringent by roughly an order of magnitude for ``homogeneous" initial conditions, for adiabatic initial conditions they were not improved nearly as much. In this paper we have chosen to remain agnostic about the initial conditions for perturbations and therefore concentrate only on the BBN bound \eqref{BBNbound}, which depends only on the number of relativistic degrees of freedom at BBN. Also, even in the case of homogeneous initial conditions, the constraints on the Hubble rate would only lower the maximum $N(k)$ by roughly one e-fold.

On the other hand, the lack of detection of a stochastic GW background places a constraint on the $(\rho_{\rm RD}, w,r)$ parameter space which can be written in terms of a lower bound on $\rho_{\rm RD}$ as follows. Judging from the steepness of the sensitivity curves of the current or future GW experiments such as LIGO or LISA compared to that of the $h^2\Omega_{\rm GW}^{(0)}(f)$ curve, as we vary the parameters $(\rho_{\rm RD}, w,r)$, the GW energy spectrum $h^2\Omega_{\rm GW}^{(0)}(f)$ first intersects with the detector sensitivity curves close to the frequencies $f_{\rm det}$ where the detectors are at their best sensitivities, $h^2\Omega_{\rm GW}^{(\rm det)}$. Hence, if a detector fails to detect a primordial GW background, the following approximate constraint can be placed \cite{Figueroa:2019paj}
\begin{equation}
    h^2\Omega_{\rm GW}^{(0)}(f_{\rm det})\lesssim h^2\Omega_{\rm GW}^{(\rm det)}\,, \label{LIGOLISAtip}
\end{equation}
where ``det" denotes different GW detectors, and the values of $f_{\rm det}$ and $\Omega_{\rm GW}^{(\rm det)}$ are listed in Table \ref{table:detectors} for different experiments. Then, using the conversion \cite{Figueroa:2019paj}
\begin{equation}
    \frac{f_{\rm RD}}{\text{Hz}}\simeq 1.5\times 10^8\left(\frac{\rho_{\rm RD}^{1/4}}{10^{16}\GeV}\right)\,,
\end{equation}
and assuming $f_{\rm RD}\ll f_{\rm det}$ and $h^2\Omega_{\rm GW,plat.}^{(0)}<h^2\Omega_{\rm GW}^{\rm (det)}$, we can rewrite Eq. \eqref{LIGOLISAtip}  as
\begin{equation}
    \ln\left(\frac{\rho_{\rm RD}^{1/4}}{10^{16}\GeV}\right)\gtrsim \Theta_{\rm det}(w,r) \label{TRDmin3}
\end{equation}
with 
\begin{equation}
    \Theta_{\rm det}(w,r)\equiv A_{\rm det}+\frac{1}{2}\left(\frac{3w+1}{3w-1}\right)\left[B_{\rm det}+\text{ln}\left(\frac{r}{0.1}\right)\right]\label{ThetaDet}
\end{equation}
and
\begin{equation}
    A_{\rm det}\equiv\ln\left(\frac{f_{\rm det}}{1.5\times 10^8\text{ Hz}}\right),\quad 
    B_{\rm det}\equiv -\ln\left(\frac{h^2\Omega_{\rm GW}^{(\rm det)}}{ 10^{-16}}\right)\,.\label{ABdet}
\end{equation}
The values of $A_{\rm det}\,, B_{\rm det}$ are listed in Table~\ref{table:detectors} for different experiments\footnote{We note that there is some variation in the numbers found in the literature. However, as long as the resulting constraints from these detectors are less stringent than the BBN bound, their exact values are not important for our purposes. This is clearly true for all the detectors listed in Table 1, apart from (the most optimistic version of) BBO for which the difference is marginal.}. However, the bound \eqref{TRDmin3} does not change our result\footnote{If $0.01\lesssim r\lesssim 0.06$, then Eq.~\eqref{TRDmin3} does not apply for BBO because then $h^2\Omega_{\rm GW,plat.}^{(0)}>h^2\Omega_{\rm GW}^{\rm (det)}$. Instead, BBO would simply rule out the aforementioned range of $r$. This would not affect the upper bound on $N$, Eq.~\eqref{NmaxBBN}, from the BBN bound since it is independent of $r$.} 
because for $w=1$ it is always the case that $\Theta_{\rm BBN}{>}\Theta_{\text{det}}$, as one can check by substituting the values in Table \ref{table:detectors} into Eq. \eqref{ThetaDet} and comparing with Eq. \eqref{ThetaBBN}.

\begin{table}
\begin{center}
\normalsize
\begin{tabular}{| c | c | c | c | c |}
\hline
Detector & $f_{\rm det}$   & $h^2\Omega_{\rm GW}^{(\rm det)}$  & $A_{\rm det}$ & $B_{\rm det}$  \\
\hline
{\rm LIGO O2} & 30 Hz & $5\times 10^{-9}$ & -15 & -18  \\ 
{\rm LIGO O5} & 30 Hz& $6\times 10^{-10}$ & -15 & -16 \\ 
{\rm LISA} & 3 mHz & $2\times 10^{-14}$ & -25 & -5 \\ 
{\rm ET} & 30 Hz& $5\times 10^{-12}$ & -15 &  -11\\ 
{\rm BBO} & {0.2 Hz} & {$1\times 10^{-17}$} & {-20} & {2} \\ 
\hline
\end{tabular}
\caption{\label{table:detectors}Optimum frequencies and best sensitivities of current and planned gravitational wave detectors, together with their $A_{\rm det}$ and $B_{\rm det}$ values as defined in Eq.~\eqref{ABdet}. Here we have computed the values for LIGO O2/O5 and LISA from the results presented in Ref. \cite{Figueroa:2019paj}, and used the limits presented in Refs. \cite{Maggiore:2019uih} (Einstein Telescope, ET) and \cite{Schmitz:2020syl} (BBO).
}
\end{center}
\end{table}

Finally, let us discuss bounds on other observables. The new bound \eqref{NmaxBBN} is only valid for $r\gtrsim 10^{-12}$, and if $r$ was smaller than this, the result \eqref{NTBBN} remains as a valid upper limit. However, as the next generation CMB B-mode polarization experiments such as BICEP3 \cite{Wu:2016hul}, LiteBIRD~\cite{Matsumura:2013aja} and the Simons Observatory \cite{Simons_Observatory} aim at detecting or constraining $r$ only at the level $\mathcal{O}(10^{-3})$, it seems unlikely that the constraint on $r$ could be improved by more than 9 orders of magnitude in any foreseeable future. Thus, we conclude that the limit \eqref{NmaxBBN} is a robust upper limit on the amount of inflationary expansion of the observable Universe between horizon exit of a scale $k$ and the end of inflation. This limit should therefore provide useful guidance in building consistent models of inflation, as they are now more constrained by our limit through the constraint on maximum $N(k)$. For more examples of this, see e.g. Refs. \cite{Martin:2010kz,Martin:2013tda, Martin:2014nya,Tenkanen:2019jiq,Allahverdi:2020bys}.


\section{Conclusions}
\label{sec:conclusions}

In this paper, we have used constraints on the number of relativistic degrees of freedom during BBN to derive a new, robust upper limit on the amount of expansion of the Universe between horizon exit of the largest observable scales today and the end of inflation, Eq. \eqref{NmaxBBN}. By comparing this result to the previous bound, Eq. \eqref{NTBBN}, one can see that for the maximum allowed value of $r$, the new bound is more stringent by $\Delta N \simeq -8$. 
The bound on the energy density of gravitational waves at the time of BBN therefore places a constraint on the amount of inflationary expansion which is stricter than any previous bound. 

We discussed the implications on inflationary models and showed how the new constraints affect model selection, and also the sensitivities of the current and planned gravitational wave observatories such as LIGO and LISA. In particular, we showed that the constraints they could impose are always less stringent than the BBN bound we obtained on $N$. We also showed that the next generation CMB B-mode polarization experiments are unlikely to improve this limit in any foreseeable future. We hope that our results will provide useful guidance in building consistent models of inflation.


\acknowledgments

We thank E. Berti, M. Kamionkowski, D.E. Kaplan, K. Kohri, S. Rajendran, T.L. Smith, T. Takahashi, V. Vaskonen, and K. Wong for correspondence and discussions. T.T. was supported by the Simons Foundation. 


\bibliography{references}
\bibliographystyle{JHEP}

\end{document}